\newcommand{\PLA}[3]{Phys.\ Lett.\ A\ {\bf #1},\ #2 (#3)}

\newcommand{\PRL}[3]{Phys.\ Rev.\ Lett.\ {\bf #1},\ #2 (#3)}

\newcommand{\PRA}[3]{Phys.\ Rev.\ A\ {\bf #1},\ #2 (#3)}

\newcommand{\diracslash}[1]{#1\llap{/\kern2pt}}

\newcommand{\be}{\begin{equation}}
\newcommand{\ee}{\end{equation}}
\newcommand{\bea}{\begin{eqnarray}}
\newcommand{\eea}{\end{eqnarray}}
\newcommand{\ba}[1]{\begin{array}{#1}}
\newcommand{\ea}{\end{array}}

\documentclass[floats,secnumarabic,nobibnotes,nofootinbib,aps,prl,showpacs]{revtex4}
\usepackage{graphicx}
\usepackage{color}
\begin{document}

\preprint{APS/123-QED}

\title{Entanglement of superposition and pair coherent states}

\author{Jitesh R. Bhatt}
\email{jeet@prl.res.in}
\author{Prasanta K. Panigrahi}

\affiliation{ Physical Research Laboratory, Navrangpura, Ahmedabad 380 009,
India}        


\begin{abstract}
 Entanglement of a pair coherent state is constructed using an iterative procedure
based on the entanglement of superpostion. It is found that only a few states
can contribute significantly to the entanglement. It is demonstrated
that the entanglement gain when a pair coherent state is superposed with
a number state can be negative.

\end{abstract}

\maketitle

Quantum entanglement is regarded to be a physical resource in quantum information
processing. Though entangled states are routinely being formed in laboratories,
there remains many open problems in classifying and understanding them(for a 
recent review see Ref. \cite{horo1}). Discrete dimensional
states are well studied and we have necessary and sufficient conditions for 
separability for $2\otimes2$ and $2\otimes3$ dimensions \cite{horo1, peres}.
Continuous variable statess are often encountered in experimental situations.
However there remains many open problems in  understanding entanglement of continuous
variable or infinite dimensional states.  
It ought to be mentioned that 
criteria for separability for such states have been obtained either using
Peres-Horodecki \cite{peres} conditions or using EPR-like operators \cite{dsw, asoka}.
But fortunately the measure of entanglement of a bipartite pure system, whether
continuous or discrete, is well understood.
The entanglement of a bipartite pure state is completely defined by
von Neumann entropy of the reduced density matrix of either of the parties \cite{bennet}.  
Thus if $\vert \Psi\rangle$ is a bipartite state shared
by two parties $A$ and $B$, its entanglement measure is defined as
\be
E(\Psi)\,=\,S(Tr_A\vert\Psi\rangle\langle\Psi\vert)=S(Tr_B\vert\Psi\rangle\langle\Psi\vert).
\ee
It is well known that  entanglement arises due to superposition of
states i.e., it is not possible, in general, to describe a composite state by
assigning  only a single state vector to any of the subsystems.
Given this basic relationship between superposition and entanglement, it is natural
to  link entanglement of some state $\vert\Gamma\rangle$ to the entanglement of the states
appearing in its superpostion expansion. It is surprising that  this
link was investigated in Ref.\cite{linden} only recently. One of the possible reasons could
be that general characterization and measures of entanglement are still not on
the firm footing \cite{horo1} and most investigations are focused only in that direction.
Since for a bipartite system these issues have been resolved, it is possible to start
investigating the link between entanglement and superposition.
In Ref.\cite{linden}, it was demonstrated that have if a bipartite state 
$\vert \Gamma\rangle$ expanded 
as a superposition of states $\vert\Psi\rangle $ and $\vert\Phi\rangle $ i.e.,
\be
\vert \Gamma\rangle\, =\,\alpha\vert\Psi\rangle +\,\beta\vert\Phi\rangle
\ee
then how $E(\Gamma)$ is related to $E(\Psi)$ and $E(\Phi)$.
This relationship is a complex one in general and depends on orthogonality relation between
$\vert\Psi\rangle $ and $\vert\Phi\rangle $. It becomes particularly simple if
$\vert\Psi\rangle $ and $\vert\Phi\rangle $ are biorthogonal\cite{linden}
\begin{eqnarray}
E(\Gamma) =|\alpha|^2 E({\Phi}) + |\beta|^2 E({\Psi}) + h(|\alpha|^2),\label{bi-orthog-equality}
\end{eqnarray}
where, $h(x)=-x\log_2 x -(1-x)\log_2(1-x)$ is the binary entropy function, and $|\alpha|^2 + |\beta|^2 = 1$. Increase in the entanglement in the superpostion
(entanglement gain) should satisfy  an upper bound \cite{linden} 
\begin{eqnarray}
E(\Gamma) -\Big(|\alpha|^2 E({\Phi}) + |\beta|^2 E({\Psi}) \Big)\leq 1,
\label{increasebound}
\end{eqnarray}
But for the cases when the superposed states are not biorthogonal eq.(3) does not satisfy.
In this case one can obtain an upper bound on $E(\Gamma)$ while the entanglement gain can become 
greater than one ebit. Given a its entanglement can be described
by eq.(1). The procedure given in Ref.\cite{linden},
provides valuable insights into how different kind of superpostions could contribute  
in constructing the entanglement of a state like $\vert\Gamma\rangle$.
 Afterwards several authors have further investigated and generalized
this result. Among the new results include different measures of entanglement \cite{measures},
multipartite entanglement \cite{multi}, including more than two states in superposition \cite{more}
and finding a new and tighter bounds \cite{Gour}.

In this work we consider an example of an entangled non-Gaussian radiation field
to analyze  the role of  entanglement of superposition.
The basic aim of this work is to analyze the role of entanglement of superpostion in
a continuous variable or infinite dimensional system.
For this purpose we choose an example of  a pair coherent state \cite{gsa_pc}.
These states  exhibit nonclassical properties\cite{tara} and they have been extensively studied
for violation of Bell inequalities \cite{gilchrist, taragsa}. Nonseparability for these states
has been established using Peres-Horodecki criterion in Ref. \cite{asoka}
A pair coherent state $|\zeta, q\rangle$ is a state of a
two-mode radiation field \cite{gsa_pc} satisfies the following properties:
\begin{eqnarray}
a b|\zeta, q\rangle =\zeta |\zeta,q\rangle\;,\\
(a^\dagger a-b^\dagger b)|\zeta,q\rangle =q|\zeta,q\rangle\;,
\end{eqnarray}
where $a$ and $b$ are the annihilation operators associated with
two modes, $\zeta$  is a complex number, and $q$ is the degeneracy
parameter. The pair coherent state for $q=0$ case, corresponding  to
equal number of photons in both the modes, is given by
\begin{equation}
|\zeta,0\rangle=N_0\sum_{n=0}^\infty
\frac{\zeta^n}{n!}|n,n\rangle\;, \label{pcs}
\end{equation}
where $N_0=1/\sqrt{I_0(2|\zeta|)}$ and $I_0(2|\zeta|)$ is the
modified Bessel function of order zero.  The density matrix can be written as
\begin{equation}
\rho=\left(\sum_{n=0}^\infty C_{nn}|n,n\rangle\right)\left(\sum_{m=0}^\infty C_{mm}^*\langle m,m|\right)\;,
\end{equation}
\noindent
where $C_{mm}=N_0\frac{\zeta^m}{m!}$ 

 A pair coherent state $|\zeta,0\rangle$ 
can be regarded as a specific kind of  superposition of biorthogonal states.
In what follows we show that one can construct the entanglement  of a pair coherent
state  using the relationship between the superpostion and entanglement provided by eq.(3).
One can also directly calculate the entanglement of the state in a 
straight forward fashion(see eq.(11) below)
which could be similar to the method given in Ref. \cite{more}. But the usage of eq.(3)
for the calculation requires an iterative procedure to superpose the infinite states in eq.(7).
However such iterative  procedure, we believe, may be useful in  providing insight 
into quantifying the relative significance of the terms in the superposition expansion
of state. This in turn can be helpful in constructing approximate states from the exact
ones.

Partial transpose of eq.(8) was shown to have negative eigenvalues
and therefore the nonseparability \cite{asoka}:
\begin{eqnarray}
\lambda_{nn}&=&\frac{1}{I_0(2|\zeta|)}\frac{|\zeta|^{2n}}{(n!)^2}\;,\;\;\;\forall n\nonumber\\
\lambda_{nm}^\pm
&=&\pm\frac{1}{I_0(2|\zeta|)}\frac{|\zeta|^{n+m}}{n!m!}\;,\;\;\forall
n\neq m\;. \label{eigenvalues}
\end{eqnarray}
One can in fact construct negativity ${\cal N}(\rho)$ by finding absolute sum of  negative eigenvalues
in lieu of a computable measure\cite{vidwer} of entanglement as,
\bea
{\cal N}(\rho) \,=\,\left\vert \sum_{n=0}^{\infty}\sum_{m=0}^{\infty}\lambda_{nm}^{-}
\right\vert=
\left\vert 1-\frac{e^{2\vert\zeta\vert}}{I_0(2\vert\zeta\vert)}\right\vert\,\, \forall n\neq m
\eea
\noindent
In the limit $\vert\zeta\vert\rightarrow\, 0$,  ${\cal N}(\rho)\rightarrow\,0$ which is indicating that
there is no entanglement. Since in this limit only $\vert\,0,0\rangle$ state will survive
in eq.(7) and all such states like $\vert n,n\rangle$ are separable. But for higher values
of $\vert\zeta\vert$ the negativity increase monotonically as depicted in Fig.(1).
\begin{figure}
\centerline{\scalebox{0.7}{\includegraphics{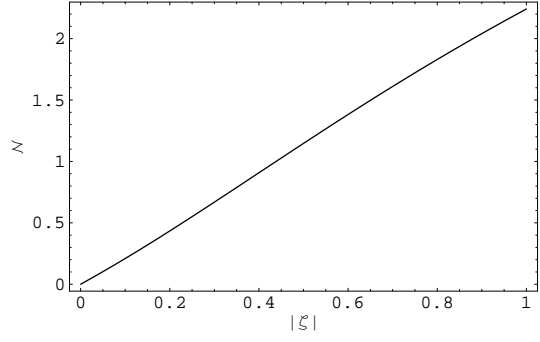}}}
\caption{\label{fig1}Variation of the negativity of the
pair coherent state  with $|\zeta|$. The
negativity monotonically increases with $|\zeta|$.}
\end{figure}

Entanglement $E(\vert \zeta,0\rangle)$ can be written as:
\begin{equation}
E(\vert \zeta,0\rangle)=-\sum_{n=0}^\infty\frac{|\zeta|^{2n}}{I_0(2|\zeta|)n!^2}\textrm{log}_2\left(\frac{|\zeta|^{2n}}{I_0(2|\zeta|)n!^2}\right)
\;. \label{entanglementpc}
\end{equation}

In order to compute the entanglement of superposition let us first consider
a set of orthogonal bipartite state $\{\vert\,i,i\rangle\}_{i=0,1,...\infty}$.
Next consider a superpostion of the first two states i.e., $N_0\{\vert\,0,0\rangle\,+\,\zeta\vert\,1,1\rangle\}$ where, $\zeta$ is a complex number.
From this, one can define a normalized state $\vert\,S1\rangle$
\begin{equation}
 \vert \,S1\rangle\,=\,N_0\gamma_{S1}(\vert\,0,0\rangle\,+\,\zeta\vert\,1,1\rangle)
\end{equation}
\noindent
where, $\gamma_{S1}=\frac{1}{N_0{\sqrt{1+\vert\zeta\vert^2}}}$. Since  $\vert S1\rangle$ 
is  expanded using two biorthogal states, the entanglement of superposition$E(S1)$ 
is written as,
\be
E(S1)\,=\, N_{0}^{2}\gamma_{S1}^{2}
\left(E(\vert\,0,0\rangle )\,+\,\vert\zeta\vert^2 E(\vert\,1,1\rangle )\right)
\,+\,h_1\left(N_0^2\gamma_{S1}^2\right).
\ee
\noindent 
As all the states in the orthonormal set that we have considered are separable, entanglement
entropies of each individual state is zero and consequently we have;
\be
E(S1)\,=\,h_1\left(N_0^2\gamma_{S1}^2\right),
\ee
which also describes the gain in the entanglement by the superposition and it also
satisfies the
condition(3). The gain always remain less than  one ebit and only in the limit $\vert\zeta\vert\rightarrow 1$,
it becomes one ebit. Now consider yet another state $\vert S2 \rangle$ defined as a superposition of states $\vert S1 \rangle$ and $\vert 2,2 \rangle$ as follows:

\be
\vert S2 \rangle=N_0\gamma_{S2}\left(\gamma_{S1}\vert S1 \rangle\,+\,\vert 2,2 \rangle\right)
\ee
\noindent
where the normalization constant is
$\gamma_{S2}\,=\,\left(N_0\sqrt{\sum_{k=0}^2\frac{\vert\zeta\vert^{2k}}{k!^{2}}}\right)^{-1}$.
The entanglement $E(S2)$ for the state defined by eq.(15)  can be written as:
\be
E(S2)=\left(\frac{\gamma_{S2}}{\gamma_{S1}}\right)^2E(S1)\,+\, h_2\left(\frac{\gamma_{S2}}{\gamma_{S1}}\right)^2
\ee
\noindent
where, $h_2$ is the gain for the second superposition carried out in defining $\vert S1\rangle$
and it can be shown to satisfy condition(3). Similarly one can construct a superposition
of the type 
\be
\vert\,Sn\rangle\,=\,N_0\gamma_{Sn}\left(\sum_{k=0}^n\frac{\zeta^k}{k!}
\vert k,k\rangle\right)
\ee
\noindent
where, the normalization constant $\gamma_{Sn}=\left(N_0\sqrt{ \sum_{k=0}^n\frac{\vert\zeta\vert^{2k}}{k!^2}} \right)^{-1}$
Entanglement $E(Sn)$ arising by superposing $\vert S_{n-1}\rangle$
and $\vert\,n,n\rangle$ state can be written as;
\be
E(Sn)\,=\,\left(\frac{\gamma_{Sn}}{\gamma_{Sn-1}}\right)^2E(Sn-1)\,+\,
h_n\left(\frac{\gamma_{Sn}}{\gamma_{Sn-1}}\right)^2
\ee
\noindent
In the limit $n\rightarrow\,\infty$ the gain in the entanglement $h_n$ becomes zero because
$\left(\frac{\gamma_{Sn}}{\gamma_{Sn-1}}\right)\rightarrow\,1$. Thus in the large $n$ limit
$E(Sn)\,\sim\,E(Sn-1)$ and the further superpositions may not contribute significantly to the entanglement characteristics of the state. $E(Sn)$ can be expressed in terms of $h_n$ as
follows:
\be
E(Sn)\,=\, \sum_{k=1}^n\left(\frac{\gamma_{Sn}}{\gamma_{Sk}}\right)^2h_k\left( \frac{\gamma_{Sk}}{\gamma_{Sk-1}}\right)^2.
\ee
\noindent
It should be noted that $E(Sn)$ can not be written as a simple sum of $n$ gains. Each time
when one superposes a new state with $\vert Sn\rangle$, a redefinition of all the previously superposed states becomes necessary.
Alternatively eq.(19), after some algebraic manipulations, can be rewritten as
\begin{equation}
E(Sn)\,=\,-\sum_{k=0}^n\,N_0^2\gamma_{Sn}^2\frac{\vert\zeta\vert^{2n}}{k!^2}log_2
\left(N_0^2\gamma_{Sn}^2\frac{\vert\zeta\vert^{2n}}{k!^2}\right)
\end{equation}
\noindent
For the limiting case when $n\rightarrow\,\infty$, the factor $N_0^2\gamma_{Sn}^2
\rightarrow\,\frac{1}{I_0(2\vert\zeta\vert)}$  and consequently $E(S_n)\rightarrow E(\vert\zeta,0\rangle)$. i.e., one obtains the entanglement of a pair coherent
state using the entanglement of superposition.
In Fig.(2), we plot $E(\vert\zeta,0\rangle)=S_a$ as a function of $\vert\zeta\vert$,
which increases with $E(\vert\zeta,0\rangle)=S_a$.  This is fine because each mode contains infinite
number of states. This plot is essentially the
same as in Ref.\cite{asoka}. Figure (3) shows two curves describing
entanglement of states $\vert S1\rangle$ and $\vert S3\rangle$.
The lower curve corresponds  to $E(S1)$, while the upper curve 
corresponds to $E(S3)$. The upper curve provides a {\it good} approximation to the entanglement
of a pair coherent state $E(\vert\zeta,0\rangle)$ shown in Fig.(1). Superposing more and more states,
with the given range of $\vert\zeta\vert$, do not alter the entanglement in a significant
way.
\begin{figure}
\centerline{\scalebox{0.7}{\includegraphics{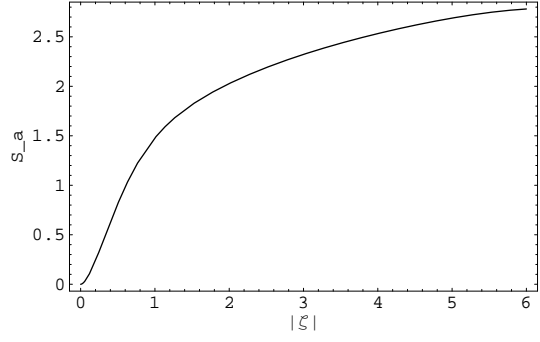}}}
\caption{\label{fig2}Variation of the von Neumann entropy(eq.(11)) of the
pair coherent state with $\vert\zeta\vert$}
\end{figure}

\noindent

\begin{figure}
\centerline{\scalebox{0.7}{\includegraphics{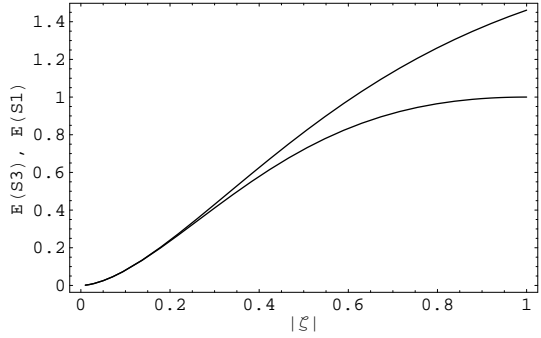}}}
\caption{\label{fig3}Variation of the entanglement $E(S3)$ and $E(S1)$ of the
pair coherent state with $\vert\zeta\vert$}
\end{figure}

Next we consider a state $\vert\Gamma_s\rangle$ which is given as a superpostion
of a pair coherent state and a number state i.e.,
\begin{equation}
\vert\Gamma_s\rangle\,=\,\alpha\vert\zeta,0\rangle\,+\,\beta\frac{\zeta^{m}}
{\sqrt{I_0}m!}
\vert\,m,m\rangle.
\end{equation}
\noindent
States $\vert\Gamma_s\rangle$ and $\vert\,m,m\rangle$ are nonorthogonal.
Density matrix for eq.(21) can be written as $\rho=\frac{1}{\Vert \Gamma_s\Vert^2}\vert\Gamma_s\rangle\langle\Gamma_s\vert$; where $\Vert \Gamma_s\Vert^2=\vert\alpha\vert^2+g^2f_m$ and $g^2=\vert\beta\vert^2+\alpha\beta^*+\beta\alpha^*$. The eigenvalues $\lambda^n$ of the
reduced density matrix $\rho^A=Tr_B(\rho)$ can be found to be
\begin{equation}
\lambda^n=\frac{\left(\vert\alpha\vert^2\,+\,g^2\delta_{mn}\right)f_n}
{\Vert \Gamma_s\Vert^2}
\end{equation}
\noindent
where , $f_n=\frac{\vert\zeta\vert^{2n}}{I_0\,n!^2}$. The integer $m$ is fixed by the number state chosen for the superposition in eq.(21) and $n=0,1,...\infty$. It is easy to check that
$\sum_{n=0}^{\infty}\lambda^n=1$ and $0\leq\lambda\leq 1$.
From the above one can write entanglement $\vert\Gamma_s\rangle$ as
\begin{equation}
E(\Gamma_s)\,=\,-\sum_{n=0}^{\infty}\lambda^nlog_2\lambda^n
\end{equation}
\noindent
It is rather easy to sum the above series but before that we make
following parameterization. Let $\vert\alpha\vert^2+\vert\beta\vert^2fm=1$,
$\alpha=cos\theta$ and $\beta=\frac{sin\theta}{\sqrt{f_m}}$. One can now
express $\Vert \Gamma_s\Vert^2=1+2cos\theta\,sin\theta\sqrt{f_m}$ and $g^2f_m
=sin^2\theta+2cos\theta\,sin\theta\sqrt{f_m}$.
Finally one can write
entanglement $E(\Gamma_s)$ for  state $\vert\Gamma_s\rangle$ as
\begin{equation}
\Vert\Gamma_s\Vert^2E(\Gamma_s)= -\Vert\Gamma_s\Vert^2log_2\frac{1}{\Vert\Gamma_s\Vert^2}
-\vert\alpha\vert^2log_2\vert\alpha\vert^2(1-f_m)+\vert\alpha\vert^2E(\vert\zeta,0\rangle)+ \vert\alpha\vert^2{\cal {F}}_1+{\cal {F}}_2
\end{equation}
\noindent
Factors ${\cal {F}}_1$ and ${\cal {F}}_2$ that we have introduced are given by,
\bea
{\cal {F}}_1\,=f_mlog_2fm
&&
{\cal {F}}_2=-\left[\left(\vert\alpha\vert^2+g^2\right)f_m\right]log_2
\left[\left(\vert\alpha\vert^2+g^2\right)f_m\right]
\eea
Gain in the entanglement $\Delta E$ for the superposition of the kind
given in eq.(21) can be defined as
\bea 
\Delta E\,=\,E(\Gamma_s)\,-\,\vert\alpha\vert^2E(\vert\zeta,0\rangle),&&
\nonumber
\eea
\noindent
as states like $\vert m,m\rangle$ are separable, $E(\vert m,m\rangle)$
is not required in the above definition.

Since $\sum_{n=0}^{\infty}\lambda^n=1$ and $0\leq\lambda\leq 1$ for 
entire parameter space, $E(\Gamma_s)$ remains positive. This is consistent
with the definition of von Neumann entropy.
In fig.(4) we have plotted $E(\Gamma_s)$ as a function of $\vert\zeta\vert$
for $m=0$ and $\theta=\pi/4$. If one compares this plot with the fig.(2)
one finds that values of $E(\Gamma_s)$ are not always larger than that
of $E(\vert\zeta,0\rangle)$. A more quantitative comparison can
come from studying the behavior of the gain. Fig.(5) shows plot of
$\Delta E$ as a function of $\vert\zeta\vert$ and the values of other
parameters same as fig.(4). It clearly shows that gain in the entanglement
become negative for certain values of $\vert\zeta\vert$. For higher
values of $\vert\zeta\vert$, $\Delta E$ first becomes zero
and increases afterwards. It must be noted that for non zero value
of $m$ also the gain can become negative. In fig.(6) we plot $\Delta E$
for $m=1$ and $\theta=\pi/4$. Again the gain becomes negative for the
higher values of  $\vert\zeta\vert$ than $m=0$ case. But further
increasing the $m$ values the entanglement gain remains positive.
It must be noted here that superpostion can reduce the entanglement
or make it to zero \cite{linden}. But the kind of superposition we are describing
is different from the illustration given in Ref. \cite{linden}. For the
present choice of parameter we have $\theta=\pi/4$.
There is no phase difference between the
 $m$-th state in the expansion of $\vert\zeta,0\rangle$ in harmonic oscillator
bases and with the number state given in eq.(21). Thus superpostion
of a pair coherent state and a number state can alter entanglement characteristic
in a complex manner. The gain in entanglement can be negative, zero or
positive depending on the values of parameter $\zeta$ takes.

\begin{figure}
\centerline{\scalebox{0.7}{\includegraphics{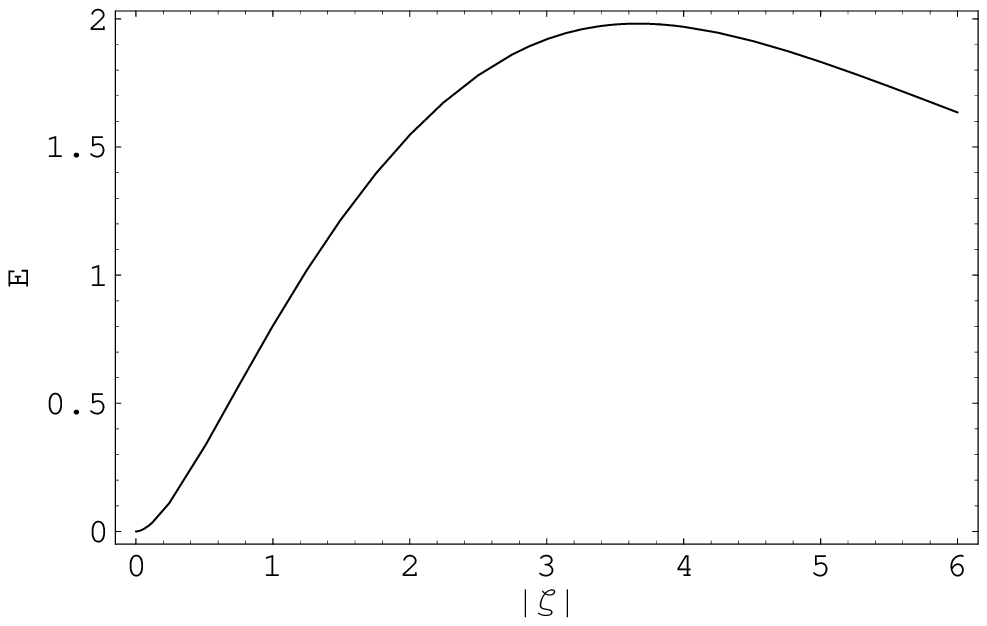}}}
\caption{\label{fig4}} The figure shows entanglement of superposition of
a pair coherent state and a number state as a function of $\vert\zeta\vert$\\ for $m=0$.
\end{figure}

\begin{figure}
\centerline{\scalebox{0.65}{\includegraphics{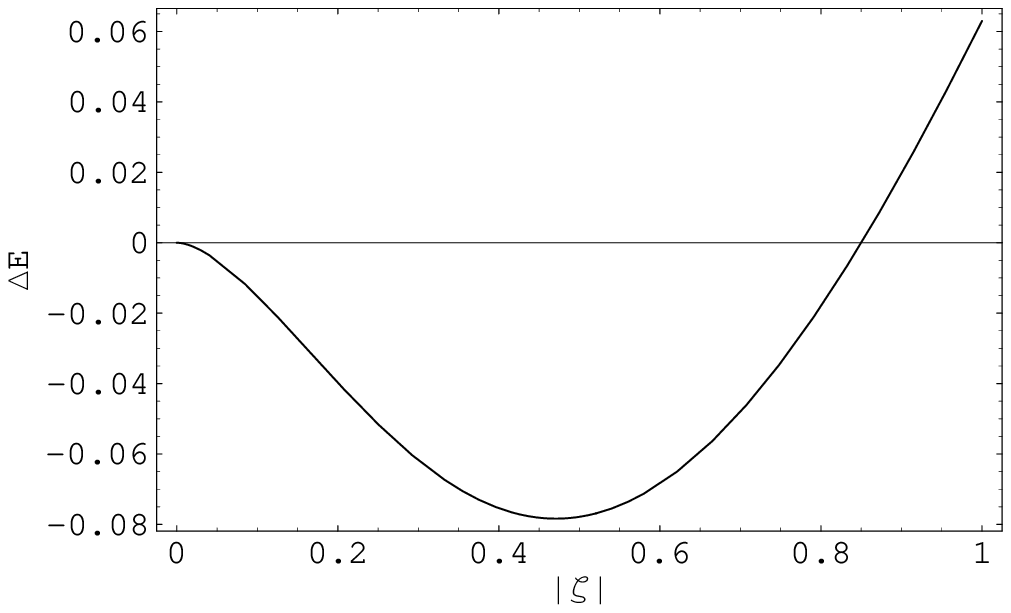}}}
\caption{\label{fig5}}
The figure shows  entanglement gain
due to the superposition a pair coherent state and a number state
as a function of $\vert\zeta\vert$ for $m=0$.
\end{figure}

\begin{figure}
\centerline{\scalebox{0.65}{\includegraphics{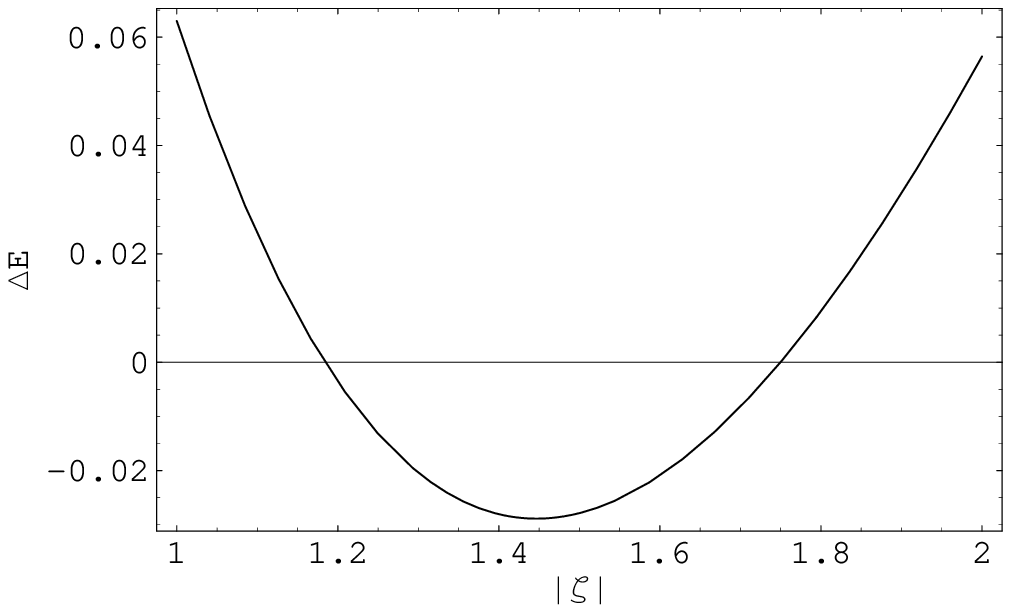}}}
\caption{\label{fig6}}The figure shows  entanglement gain
due to the superposition a pair coherent state and a number state
as a function of $\vert\zeta\vert$ for $m=1$.
\end{figure}
In conclusion, we have constructed the entanglement of pair coherent state by
considering successive gain arising due to superpostion of the number states.
We have shown that only a few states in the superposition expansion can contribute
significantly, this can be qualitatively expected from the expansion in eq.(7).
However, the method used in analysing the entanglement of superposition can provide
a better way of approximating the terms. We have shown that for a certain
parameter space (as shown in fig.(5))  the gain in entanglement  can become
negative when superposing a pair coherent state with a number space.

\end{document}